\documentstyle[12pt]{article}
\newcommand{\be}{\begin{equation}}
\newcommand{\ee}{\end{equation}}
\newcommand{\ba}{\begin{eqnarray}}
\newcommand{\ea}{\end{eqnarray}}
\newcommand{\ft}{\footnote}
\newcommand{\nl}{\newline}
\newcommand{\al}{\alpha}
\newcommand{\bt}{\beta}
\newcommand{\ga}{\gamma}

\newcommand{\la}{\lambda}
\newcommand{\Tr}{{\rm Tr}}

\begin{document}

\begin{flushright}
hep-th/9705138\\
QMW-PH-97-17\\
IC/97/46\\
(revised 3 June 1997)
\end{flushright}

\begin{center}

{\Large{\bf Higher Dimensional Analogues of Donaldson-Witten 
Theory.}

B.S.Acharya$^{*}$\ft{r.acharya@qmw.ac.uk},
M. O'Loughlin$^{\#}$\ft{mjol@ictp.trieste.it},
B. Spence$^{*}$\ft{b.spence@qmw.ac.uk \nl BSA is supported by
a PPARC Postgraduate Fellowship and
BS by an EPSRC Advanced Fellowship.}}

$^*${\it Department of Physics, Queen Mary and 
Westfield College, Mile End Road,
London E1 4NS UK.}

$^\#${\it International Center for Theoretical Physics, Trieste, 
Italy.}

\end{center}

\begin{abstract}
We present a Donaldson-Witten type field theory in
eight dimensions on manifolds with $Spin(7)$ holonomy.
We prove that the stress tensor is BRST exact 
for metric variations preserving
the holonomy and we give the invariants for this
class of variations. In six and seven dimensions we propose similar
theories on Calabi-Yau threefolds and manifolds of $G_2$ holonomy 
respectively.
We point out that these theories arise by considering 
supersymmetric
Yang-Mills theory defined on such manifolds. The theories are
invariant under metric variations preserving
the holonomy structure 
without the need for twisting. This statement is a 
higher
dimensional analogue of the fact that Donaldson-Witten field 
theory on
hyper-K\"ahler 4-manifolds is topological without twisting. 
Higher dimensional analogues of Floer cohomology are briefly 
outlined.
All of these theories arise naturally within the context 
of string theory. 
\end{abstract}
\newpage
\section{Introduction.}

Instantons in four dimensions have proven to be remarkably
important
objects for our understanding of physics and of mathematics. This
fact was reflected in \cite{Wit}, where it was shown how quantum 
field
theory could be used to construct the Donaldson invariants for
$4$-manifolds. As is well known, instantons play a key role in 
this
story. 

In \cite{cdfn} instanton equations in higher dimensional flat
space were written
down. These were subsequently discussed in \cite{bm} where we 
showed
that these equations are very naturally associated with the list 
of
holonomy groups of Ricci flat manifolds (in dimension eight or 
less).

It is natural to ask whether or not some analogue of the 
topological
quantum field theory of \cite{Wit} exists on these higher 
dimensional
manifolds. That is the purpose of this paper. In fact the theories 
we
discuss here are an extension of the Donaldson-Witten
theory \cite{Wit} to higher dimensions.
This is due to the two basic facts that $(i)$ the
correlation functions are invariant under metric deformations
which preserve the holonomy structure of the manifold
and $(ii)$ the higher dimensional instantons are  minima
of
the Yang-Mills action and can therefore be used to evaluate the 
observables
of the theory. 

In the next section we will discuss the instanton equations 
themselves, 
and review the relation with certain Ricci flat manifolds. We also 
introduce
some notation which will prove useful later in the paper. In 
section 3 we
introduce the action and show that the observables of
the theory are invariant under metric deformations
which preserve the holonomy. 
In section 4 we describe
the variational calculus (in the case of the $8$-dimensional 
theory) which
is required to verify the formulae of section 3. This calculation 
also 
proves useful from another point of view, since we are able to 
present a
very simple (local) discussion of some theorems and results given in 
\cite{J3,gibb,Wang}.
In section 5 we give explicit expressions for the BRST invariant 
observables
from which correlation functions can be constructed.
In section 6 we outline the higher dimensional version of Floer theory 
which,
as in \cite{Wit}, arises very naturally in the Hamiltonian 
formulation.

In our concluding section we point out that for the
manifolds we have been discussing these field
theories are {\it not} twisted versions of super-Yang-Mills 
theory. Rather
these theories are simply super-Yang-Mills theories formulated on 
manifolds
with reduced holonomy groups. The theories we discuss here 
are invariant under a certain class of metric variations,
however they arise
without the need for twisting. This statement is in fact a 
generalisation 
of the statement that the four-dimensional theory of \cite{Wit}
formulated
on a hyper-K\"{a}hler 4-manifold is topological {\it without} the need
for twisting
\cite{vw}.

\section{Instanton Equations in $D > 4$.}

In this section we give a brief review of the instanton equations
in $D>4$. This will set the notation and conventions for the 
following
sections.

Higher dimensional instanton equations were first written down in
\cite{cdfn}, for $4\leq D\leq 8$. As in four dimensions, these 
equations
are first order self-duality equations for gauge fields. The 
general
form of the equations is:
\be
\lambda F_{\mu\nu} =  {1\over 2}
\phi_{\mu\nu\rho\sigma}F^{\rho\sigma}
\ee
This set of equations will be the focus of this paper. In $(1)$, 
$F$ is the field strength for the gauge field $A$, and the indices run 
from
$1$ to $D$. In \cite{cdfn} these equations were considered in flat
space. However, it is natural to consider these equations on
curved manifolds \cite{bm}.
We assume that the manifold on which the gauge field propagates
is a $D$-dimensional Riemannian manifold. This means that its 
holonomy
is contained within $SO(D)$. The totally antisymmetric tensor 
$\phi$ is
a singlet of the holonomy group $H$ $\subset$ $SO(D)$. This can
only occur if $H$ is a proper subgroup of $SO(D)$ and not $SO(D)$ 
itself.
Thus, requiring that the instanton equations are non-trivial 
automatically
implies a reduction of the Lorentz group of the theory from 
$SO(D)$ to $H$.

It is a problem in group theory to calculate when the instanton 
equations
are non-trivial. This was done in \cite{cdfn}. It was further 
pointed out in
\cite{bm} that if F is the Riemannian curvature 2-form for the 
$D$-manifold,
then the solutions to $(1)$ are precisely the Ricci flat manifolds 
whose
holonomy is given by Berger's classification \cite{berger}. In 
fact for
$4 < D \leq 8$, all Ricci flat manifolds with holonomy a
proper subgroup of $SO(D)$ admit a covariantly constant 
$4$-form,
and $\phi$ is this $4$-form.
Thus the instanton equations have the possibility of being 
non-trivial
on any $4 < D \leq 8$ dimensional Ricci flat manifold,
with $\phi_{\mu\nu\rho\sigma}$ being given 
by the components of the corresponding
holonomy singlet $4$-form. We will mainly be interested in the 
cases when
$D=  8,7,6$ and $H$ is $Spin(7),{G_2},SU(3)$ respectively.

For $D= 8$ and $Spin(7)$ holonomy, $\phi$ is a Hodge self-dual 
$4$-form (for a given choice of orientation for the $8$-manifold
$M_8$). The 
instanton equations
are then non-trivial when $\lambda$ =  $-1$ and $3$. The 
components of the
tensor $\phi$ are closely related to the structure constants of 
the
octonions, and for a given choice of octonionic structure 
constants, there
exists one choice for $\phi$ (for precise details on how this 
works
see \cite{gun}). We will choose $\phi$ as in \cite{J3} and its 
components 
in an orthonormal frame are
given by:
\ba
&[1256] =  [1278] =  [3456] =  [3478] =  [1357] =  
[2468] =  [1234] =  [5678] =  1\nonumber\\
&[1368] =  [2457] =  [1458] =  [1467] =  [2358] =  [2367] = 
-1
\ea
where $[ijkl]$ means ${\phi}_{ijkl}$ and all other components are 
zero. Since this form is $Spin(7)$ invariant it induces a metric on
the $8$-manifold $M_8$. Moreover, if the form is covariantly constant
then the holonomy group of the associated metric is $Spin(7)$ \cite{bo}.
In fact this is so if and only if $\phi$ is closed \cite{brs}. Following \cite{J3}, we will refer
to $4$-forms admitting some isomorphism with $\phi$ above as {\it admissable}
$Spin(7)$ structures.

Group theoretically, the instanton equations tell us that F 
transforms
under a certain representation of the holonomy group. Precisely 
which
representation is determined by $\lambda$. For the case of $H$ =  
$Spin(7)$
holonomy, $\lambda =  -1$ puts F in a $\bf 21$ of H, and $\la =  
3$ gives
F as a $\bf 7$ of $H$. This is due to the splitting of the adjoint
of $SO(8)$ under $Spin(7)$:
\be
\bf 28 \rightarrow 21 + 7
\ee
This splitting means that ${{\Lambda}^2}(TM)$ decomposes into two 
orthogonal subspaces of $2$-forms. These subspaces consist of $2$-forms with 
$7$ and
$21$ independent components respectively. This means that we can 
introduce
projection operators which project onto the 
${\bf 7}$ and ${\bf 21}$ dimensional pieces
of any 2-form. These are given by
\be
P_7 =  {1\over 4}(1+ {1\over 2}\phi), \qquad 
P_{21} =  {3\over 4}(1 - {1\over 6}\phi).
\ee
We are using a shorthand matrix notation where, for example,
the first expression above represents
\be
{P_7}_{\mu\nu}^{\hphantom{\mu\nu}\alpha\beta} =   {1\over4}\left(
\delta^\alpha_{[\mu}\delta^\beta_{\nu]}
+ {1\over 2}\phi_{\mu\nu}^{\hphantom{\mu\nu}\alpha\beta}\right).
\ee
These projectors satisfy the relations
\ba
& (P_7)^2 =  P_7,\qquad (P_{21})^2 =  P_{21}\nonumber \\
& P_7P_{21} =  P_{21}P_7 =  0,
\qquad P_7 + P_{21} =  1.
\ea
Note that the form $\phi$ satisfies
\be
\phi^2 =  4\phi + 12.
\ee


\subsection{D= 7,6}

By setting to zero those components of $F$ and $\phi$ which 
contain (say) an
$8$ index, one gets a set of instanton equations in $D= 7$. This 
corresponds
to the reduction of $Spin(7)$ to $G_2$. Of course, the resulting 
$\phi$ is
a $G_2$ singlet. This means we are now considering the equations 
on a
$7$-manifold of $G_2$ holonomy. The instanton equations $(1)$ now 
correspond
to the restriction of $F$ to $\bf 14$ and $\bf 7$ dimensional 
representations
of $G_2$ according to the splitting of the $\bf 21$ of $SO(7)$ 
into a $\bf 14$
and $\bf 7$. In this case, $\phi$ satisfies
\be
{\phi}^2 =  8 + 2\phi.
\ee

This gives the values of $\lambda$ in $(1)$ as $-1$ and $2$, 
corresponding
to $F$ being in the $\bf 14$ and $\bf 7$ respectively.
The normalised projection operators in this case are:
\be
P_7 =  {1\over3}\left(1 + {1\over2}\phi\right),
\ee
and
\be
P_{14} =  {2\over3}\left(1 - {1\over4}\phi\right).
\ee

By setting to zero one further set of components of $F$ and 
$\phi$, which
contain (say) a $7$-index, one gets a set of instanton equations 
in $D= 6$. This corresponds to the reduction of the holonomy group 
from $G_2$ to $SU(3)$. In this case, $\phi$ is proportional to $k\wedge 
k$, where $k$
is the Kahler form of the Calabi-Yau threefold. 


\section{The Field Theories}

The purpose of this paper is to describe how the moduli space of 
solutions
to $(1)$ can be used to extract quantities (observables)
on manifolds with holonomy $H$ (as above) which are invariant 
under metric deformations
which preserve the holonomy.
These will follow from a simple generalisation 
of the field theory in \cite{Wit}, which provided 
a physical formulation of Donaldson theory.
We will construct theories associated with the moduli
space of solutions to $(1)$ using the paradigm of fields, 
symmetries and
equations \cite{wit2,geo}. This construction will turn out to have
a similar structure as that of \cite{Wit}.

As explained in \cite{Wit} and reviewed
in more detail in \cite{geo}, in topological field theories with 
gauge
symmetries one introduces an anticommuting $1$-form which is the 
fermionic
partner of the gauge field. Under the BRST symmetry the gauge 
field transforms
into this $1$-form. One further introduces a scalar field which is 
invariant
under the BRST symmetry $Q$. This scalar, being BRST invariant can 
then
be used to construct topological observables. All fields transform 
in the
adjoint of the gauge group.
With this ``multiplet'', if the fermionic 
$1$-form transforms under $Q$ into a gauge transformation 
generated by a 
parameter which is the scalar field, then $Q^2$ $= 0$, up to 
gauge 
transformations. Since ${Q^2}= 0$, one can study $Q$-cohomology. 
The $Q$
cohomology classes are the observables of the theory. Essentially, 
this
multiplet is associated with the symmetries of the theory.

In addition to the above multiplet one also introduces fields 
which encode
the moduli problem one is interested in. In the case we are 
interested in,
this includes a field which transforms under $Q$ into the 
instanton 
equation itself.
For example, in $D= 8,7$, the instanton equation asserts that 
${P_7}F= 0$ with
the appropriate $P_7$ defined above. We thus include a fermionic 
$2$-form
which transforms into ${P_7}F$. Finally, it is also useful 
to introduce two more scalars, with
opposite statistics. An action with the above structure in four 
dimensions
was given in \cite{Wit}, and it turns out that
a suitable choice of Lagrangian in higher
dimensions is provided by a similar Lagrangian to \cite{Wit}, but 
now
considered as a theory defined on a $D$-dimensional manifold. Of
course, in order to define the theory on such a manifold one 
requires
the existence of a $4$-form which is a singlet of the holonomy 
group $H$.
This group must be a proper subgroup of $SO(D)$.
In eight and lower dimensions the maximal proper subgroups are
$Spin(7)$, $G_2$ and $SU(3)$ for $D= 8,7$ and $6$.

We will give the Lagrangians for the $8$ and $7$ dimensional
cases explicitly. We expect the $6$ dimensional model to
take a similar form, with the complex structure playing an important role.
In $D= 8,7$ dimensions we will consider a theory defined on 
$M_D$
(where $M_D$ has holonomy $Spin(7)$ or $G_2$ 
respectively).
The field content in these theories will be the same as that in 
the four
dimensional theory in \cite{Wit}, except that the duality conditions
on the 2-form will be dimension-dependent relations.

Explicitly, the action for the $8$ dimensional theory is given by
\be
S =  S_1 + S_2,
\ee
where
\ba
S_1 &= & \Tr{\int_{M_8}}{d^{8}x}{\sqrt g}
\{ {1\over 4} F_{\al\bt}F^{\al\bt} + 
{1\over 2}{\varphi}D^\al D_\al{\lambda} -i{\eta}D^\al \psi_\al
 \nonumber\\ 
&+& 
2i(D_\al{\psi_\bt}).{\chi^{\al\bt}} -{i\over 4}
{\varphi}[{\chi_{\al\bt}},{\chi^{\al\bt}}]
-{i\over 2}{\lambda}[{\psi_{\al}},{\psi^{\al}}] \nonumber\\ 
&-& {i\over 2}{\varphi}[{\eta},{\eta}]
-{1\over8}{[{\varphi},{\lambda}]}{^2} \},
\ea
and
\be
S_2 =  {1\over4}\Tr{\int_{M_8}}{d^{8}x}{\sqrt
g}F_{\al\bt}\tilde{F}^{\al\bt},
\ee
where
\be
\tilde{F}_{\al\bt} =
{1\over2}{\phi}_{\al\bt\ga\delta}F^{\ga\delta}.
\ee

The action $S_2$ is the $8$ dimensional instanton action 
\cite{bm}\ft{
In \cite{bm} we stated that $S_2$ is a topological invariant.
However our later formulae show that this is not the case.}. 
In $S_1$ the 
commuting fields are the gauge field $A_\al$ for which $F_{\al\bt}$ is the
curvature, and two scalar fields, $\varphi$ and $\lambda$. The 
anticommuting
fields are a $1$-form ${\psi_{\al}}$, a self-dual $2$-form $\chi_{\al\bt}$ 
(which in the
$Spin(7)$ and $G_2$ holonomy cases is in the $\bf 7$ of the 
holonomy
group) and a scalar, $\eta$. Note that the $\varphi$ which appears 
in $S_1$ is a scalar
field and should not be confused with the $Spin(7)$ structure 
($\phi$)
which appears
in the instanton equations and in $S_2$.

The action $S$ 
is invariant under the 
BRST transformations $\delta = -i\epsilon\{Q,\hphantom{P}\}$, 
with
anticommuting parameter $\epsilon$, \be
\{Q,A\} =  - \psi
\ee
\be
\{Q,\varphi\} =  0
\ee
\be
\{Q,\lambda\} =  - 2\eta
\ee
\be
\{Q,\eta\} =  {i\over2}[{\varphi},{\lambda}]
\ee
\be
\{Q,\psi\} =  - iD\varphi
\ee
\be
\{Q,{\chi_{\al\bt}}\} =  {1\over 2}i(F + \tilde{F})_{\al\bt}.
\ee

In the $7$ dimensional case, the action is again $S_1+S_2$ as above,
with the following differences:
Firstly,  one integrates over the $7$ dimensional manifold, and utilises
the $4$-form $\phi$ appropriate to the $G_2$ case; secondly, the 
coefficients of the $(D_\al{\psi_\bt}).{\chi^{\al\bt}}$ and $
{\varphi}[{\chi_{\al\bt}},{\chi^{\al\bt}}]$ terms are
${3i\over2}$ and ${-3i\over16}$ respectively; finally,
the BRST symmetry is again of the form as in the $8$ dimensional case, except
that the $\chi$ variation is now 
$\{Q,{\chi_{\al\bt}}\} =  {2\over 3}i(F + \tilde{F})_{\al\bt}$.

Apart from factors these actions are of the same form as that in 
\cite{Wit},
and similarily have an additive ``ghost'' number symmetry ($U$), for
which the
charge assignments are $(0,2,-2,-1,1,-1)$
for the fields $(A,\varphi,\la,\eta,\psi,\chi)$, respectively.

The Lagrangian for the theories is BRST exact: 
\be
L =  -i\{Q,V\},
\ee
where $V$ is given in the $8$ dimensional case by
\be
V = {1\over 2}\Tr\left({F_{\al\bt}}{\chi^{\al\bt}} + 
{\psi_{\al}}{D^{\al}}{\lambda}
- {1\over 2}{\eta}[{\varphi},{\lambda}]\right),
\ee
and in the $7$ dimensional case by
\be
V = {1\over 2}\Tr\left({3\over4}{F_{\al\bt}}{\chi^{\al\bt}} + 
{\psi_{\al}}{D^{\al}}{\lambda}
- {1\over 2}{\eta}[{\varphi},{\lambda}]\right).
\ee


\section{On the Metric Dependence of the Theories.}

In order to prove that the correlation functions of a field theory 
on 
a manifold $M$ are independent of the metric tensor on $M$ one 
shows that the
energy momentum tensor is $Q$-exact. This can be done explicitly 
for the theories we are discussing here  if we restrict our 
attention
to manifolds of reduced holonomy and metric deformations which
preserve this.
Under a small change $\delta g_{\al\bt}$
in the metric $g$ on $M_D$, the
action changes by 
\be
{\delta}S =  {1\over 2}{\int_{M_D}}
{\sqrt g}{\delta}{g^{\al\bt}}{T_{\al\bt}},
\ee
which defines the energy-momentum tensor $T_{\al\bt}$. 

In the 
eight-dimensional case, when $M_8$ has $Spin(7)$ holonomy, 
the energy-momentum tensor is given by the following expression:
\ba
{T_{\al\bt}} &= & \Tr \{ F^\mu_{\hphantom{\mu}(\al} F_{\beta)\mu} 
+  F^\mu_{\hphantom{\mu}(\al}\tilde{F}_{\beta)\mu}
        -{1\over 4}g_{\alpha\beta}
         F_{\gamma\delta}(F^{\gamma\delta}
      +   \tilde{F}^{\gamma\delta})                       
\nonumber\\
         &+& 2i 
D_{[\alpha}\psi_{\sigma]}\chi_\beta^{\hphantom{\bt}\sigma}
+ 2i D_{[\beta}\psi_{\sigma]}\chi_\alpha^{\hphantom{\bt}\sigma}
-ig_{\alpha\beta}D_{\tau}\psi_{\sigma}\chi^{\tau\sigma}\nonumber\\
      &-&D_{(\alpha}\varphi D_{\beta)}\lambda
        +{1\over2}g_{\alpha\beta}D_{\sigma}\varphi 
D^{\sigma}\lambda
    +2iD_{(\alpha}\eta\,\psi_{\beta)}
   - ig_{\alpha\beta}D_{\sigma}\eta\;\psi^{\sigma}       
\nonumber\\
  &-&2i \lambda\psi_\alpha\psi_\beta +
ig_{\alpha\beta}\lambda\psi_\sigma\psi^\sigma
  + {i\over 2}g_{\alpha\beta}\varphi[\eta,\eta] + 
   {1\over 8}g_{\alpha\beta}[\varphi,\lambda]^2 \}.
\ea
To derive this expression for $T_{\al\bt}$, one needs to know the
variations of the holonomy structure $\phi$ and the self-dual 
2-form $\chi$, as their definitions are metric dependent. We will 
discuss this in some detail later in this section.

One can check that this energy-momentum tensor is $Q$-exact:
\be
T_{\al\bt} =  -i\{Q,\la_{\al\bt}\},
\ee
where
\be
\lambda_{\al\bt} =  \Tr\{ 2 F_{(\al}^{\hphantom{\mu}\mu}
                                   \chi_{\bt)\mu} - 
{1\over2}g_{\al\bt}
F_{\ga\delta}\chi^{\ga\delta} + 
                           \psi_{(\al}D_{\bt)}\lambda - 
{1\over2}g_{\al\bt}
                    \psi_\sigma D^\sigma\lambda + 
{1\over4}g_{\al\bt}
                       \eta[\varphi,\lambda] \}.
\ee

This of course is the key property of the theory, since it implies
that
correlation functions of BRST invariant observables are 
independent of suitable variations of the
 metric tensor on $M_8$. There are some 
key points
which we would like to clarify here. The existence of the singlet 
$4$-form
induces a metric tensor on $M_8$ (see \cite{J3} and the next 
subsection).
This metric tensor is Ricci flat and the above energy-momentum 
tensor is that
which comes from varying the Ricci flat metric associated with the 
$4$-form
$\phi$. We will show that
the correlation functions of $Q$-invariant observables are 
invariant under metric deformations
which preserve the holonomy of $M_8$. This can be done by showing 
that 
the energy-momentum tensor is $Q$-exact for these variations.

We will now give the details of the variational
calculations required to derive the above
energy-momentum tensor. We must first
compute the variation of $\phi$ under variations of the 
metric.


\subsection{Metric Dependence of the Holonomy Structure.}

In verifying the above results, one needs to note that 
the field $\chi_{\al\bt}$ satisfies a self-duality condition
involving the $Spin(7)$ structure $\phi$ and the metric tensor
$g_{\al\bt}$. The form $\phi$ itself also has a metric
dependence as we will see below. 

The $4$-index tensor $\phi$ is a singlet of a subgroup $H$ of 
$GL(D)$.
This implies that the holonomy group of the manifold 
on
which $\phi$ is defined is at most $H$. In the cases in which we are 
interested
$H$ $\subset$ $SO(D)$ and thus the manifold $M_D$ is Riemannian. 
The
existence of $\phi$ therefore induces a metric on $M_D$. 
Furthermore, every choice
of $\phi$ generically induces a different metric on $M_D$. This 
means
that given a choice of $\phi$ (and therefore a metric $g$), a 
change in
the metric induces a change in $\phi$. In this section we will
explicitly calculate what this change is. We will restrict our 
attention to the case in which $D= 8$ and $H$ is $Spin(7)$.

In our conventions $\phi$ is
Hodge self-dual:
\be
\phi =  *\phi.
\ee
With respect to the original metric $g$, there exists a splitting 
of
the space of $4$-forms on $M_8$ into the (orthogonal) spaces of 
Hodge self-dual
and anti-self dual $4$-forms. Varying the metric tensor $g$ in the 
above
equation gives a variation of $\phi$. We may therefore introduce 
projection
operators which project any $4$-form onto its self or anti 
self-dual 
components. The equation that $\phi$ is self-dual may be
written as 
\be
{P_+}\phi =  \phi
\ee
Varying this equation one obtains
\be
\delta{P_+}.\phi + {P_+}.\delta\phi =  \delta\phi
\ee
Thus
\be
(1 - {P_+})\delta\phi =  \delta{P_+}.\phi
\ee
However, 
\be
{P_+} + {P_-} =  1
\ee
and thus the variation of the $\phi$ self-duality condition yields
\be
{P_-}\delta\phi =  \delta{P_+}.\phi
\ee

In other words, the self-duality condition allows one to determine
the component of $\phi$ which varies into the anti-self-dual 
chamber.
In components one obtains that
\be
[(\delta{P_+}).\phi]_{\al\bt\ga\delta} = -{1\over
4}(\delta{g^{\sigma\tau}}g_{\sigma\tau}){\phi}_{\al\bt\ga\delta}+
2\delta{g_{\lambda[\al}}{\phi}^{\lambda}_{\hphantom{m}\bt\ga\delta]}
\ee
By the above, this expression is Hodge anti self-dual. 
However, the
first term on the right is self-dual because it is 
proportional to $\phi$. Since
we have considered arbitrary variations in the metric tensor, we 
may certainly
consider those which are traceless (which would remove this term 
from
the above expression). However, upon further consideration one 
sees that if
one writes the metric variation of the second term on the right
as the sum of a trace and traceless part, then
the trace term exactly cancels the first term. Thus, the 
expression above
is Hodge anti self-dual as required with arbitrary metric 
variations.
Thus the anti self-dual variation of $\phi$ under arbitrary 
variations of the
metric is
\be
{P_-}(\delta\phi)_{\al\bt\ga\delta} =  
2{\delta}{\tilde 
g}_{\lambda[\al}{\phi^{\lambda}_{\hphantom{m}\bt\ga\delta]}}.
\ee
where ${\delta}{\tilde g}_{\al\bt}$ is traceless.

Usually, in variational problems of this
type in which one has a decomposition of a vector space into two 
(or more)
orthogonal subspaces, in the variation with respect to the
metric of a vector which belongs 
solely to one of the subspaces, one can only determine the 
variation of the
vector into the orthogonal subspace. The orthogonal component of 
the variation
can then only be determined if the original vector satisfies 
further identities
which involve the metric tensor. This is the case for $\phi$.

In flat space, $\phi$ satisfies identities which are given in 
\cite{gun}.
These are easily generalised to curved spaces by replacing the 
flat
space metric by the curved one. 
As these identities are crucial in proving the relations given in 
this paper,
we reproduce them here. 
The basic one is
\be
\phi_{\mu\al\bt\gamma}\phi^{\mu\nu\rho\sigma} = 
    6\delta^\nu_{[\al} \delta^\rho_\bt \delta^\sigma_{\gamma]}
+ 
9\phi_{[\al\bt}^{\hphantom{\al\bt}[\nu\rho}\delta_{\gamma]}^{\sigma]
},
\ee
from which by contractions one deduces that
\be
\phi_{\mu\nu\bt\gamma}\phi^{\mu\nu\rho\sigma} = 
    12\delta^\rho_{[\bt} \delta^\sigma_{\gamma]}
+   4\phi_{\bt\gamma}^{\hphantom{\al\bt}\rho\sigma},
\ee
\be
\phi_{\al\mu\nu\rho}\phi^{\hphantom{\bt}\mu\nu\rho}_{\bt} 
=  42g_{\al\bt},
\ee
and
\be
\phi_{\mu\nu\rho\sigma}\phi^{\mu\nu\rho\sigma} =  42.8
\ee
The factors in these expressions are the secret of
the consistency of our results.
It is important to note also
that 
the variation (35) can be checked to be consistent with these
relations.

Having calculated the Hodge anti self-dual variation of $\phi$, we 
still
require calculating the self-dual variation.
In order to proceed with this, one
now needs to compare what we have learned above with what is known
about the moduli space of $Spin(7)$ structures on $M$. The space 
of all
possible choices for $\phi$ (modulo diffeomorphisms isotopic to 
the
identity) was studied in \cite{J3}. In particular, it was shown
that this moduli space is $b_4^-$ $+$ $1$ dimensional, where 
$b_4^-$ is the
number of harmonic Hodge anti self-dual four-forms on $M$. 
Further, the moduli space
of Ricci flat metrics 
was studied in \cite{gibb}. This space is also $b_4^-$$+1$
dimensional. One would therefore expect that these two spaces are 
isomorphic.  
To show this locally let us first compare what we have learned above to
the considerations in \cite{gibb}. In the above we have shown 
that the anti-self-dual
variation of $\phi$ with respect to
 the metric tensor of $M$ is given by:
\be
(\delta\phi)_{\al\bt\ga\delta} 
=  2{\delta}{\tilde g}_{\lambda[\al}{\phi^{\lambda}_{\hphantom{\lambda}
\bt\ga\delta]}}
\ee
where the variation of the metric in the above is restricted to be 
traceless.
In \cite{gibb} it was shown that given a symmetric, 
{\it traceless} $2$-index object $h_{\lambda\al}$ that the 
$4$-index tensor
\be
h_{\lambda[\al}{\phi^{\lambda}_{\hphantom{\lambda}\bt\ga\delta]}}
\ee
is Hodge anti self-dual\ft{The orientation conventions of 
\cite{gibb} are
different from ours.}.

But this is precisely the form of our above expression for the 
anti self-dual
variation of the $Spin(7)$ structure $\phi$. Furthermore it was 
shown in
\cite{gibb} that the above mapping between anti self-dual 
$4$-forms and
symmetric traceless $2$-index tensors is invertible and 
$1$-to-$1$. Thus
the space of $4$-forms is isomorphic to the space of symmetric, 
traceless
$2$-index tensors. It is also true that if the
anti self-dual $4$-form above is harmonic, then the traceless 
and divergenceless $\delta {\tilde g}_{\al\bt}$ is a zero-mode of the 
Lichnerowicz equation \cite{gibb}.
This means that changing the $Spin(7)$ structure on the
manifold by adding to it a small harmonic anti self-dual $4$-form is 
{\it equivalent}
to volume preserving Ricci flat metric deformations. 

The key question now is whether or not $\phi^{\prime}$ $=$ $\phi$ $+$
$\delta\phi$ is a torsion free $Spin(7)$ structure associated with the
new Ricci flat metric $g^{\prime}$ = $g$ $+$ $\delta g$. One can show that
it is in two ways. First, since our first order variations are compatible
with the identities satisfied by $\phi$, $\phi^{\prime}$ also
satisfies these identities. But these are precisely the identities
satisfied by an arbitrary $Spin(7)$ invariant $4$-form. Since the result
of \cite{gibb} implies that when $g^{\prime}$ is Ricci flat $\phi^{\prime}$
is closed, we may conclude by theorem 3 of \cite{brs} that $\phi^{\prime}$
is a torsion free $Spin(7)$ structure. Thus the infinitesimal volume
preserving Ricci flat metric variations preserve the holonomy group.

Another way to arrive at the same result is by using the construction
of $\phi$ by parallel spinors \cite{Wang}. In the second reference of
\cite{Wang} it is shown that under infinitesimal Ricci flat metric
variations, a parallel spinor always exists. This means that a covariantly
constant, $Spin(7)$ invariant $4$-form can always be constructed in the
nearby Ricci flat metric.

Apart from these
deformations one may also make a scaling of
the metric i.e. change the volume. 
Since
we know that the moduli space of $Spin(7)$ structures (modulo 
diffeomorphisms)
has dimension $b_4^-$ $+$ $1$ \cite{J3}
and our first order formula for the anti self-dual 
variation of $\phi$ has allowed us to identify $b_4^-$ of
these variations, we expect that $\phi$ 
must scale
if we scale the metric tensor. If this were not so the extra 
coordinate
on the moduli space of $Spin(7)$ structures would be missing and 
our moduli
space would only have dimension $b_4^-$. In general under a 
transformation of
the metric whose traceless part vanishes $\phi$ will change as
follows:
\be
\delta\phi =  k(\delta g^{\al\bt}g_{\al\bt})\phi
\ee

In order to determine the constant $k$, one varies the identities 
which
are satisfied by $\phi$.
This then  fixes $k$ =  $-1/4$. Adding this part of the 
variation of
$\phi$ to its anti self-dual variation given in $(35)$ one finds 
that under
an arbitrary change in the metric, the change in $\phi$ is given 
by
\be
(\delta\phi)_{\al\bt\ga\delta} =  
2\delta 
{g_{[\al\lambda}}{\phi^{\lambda}_{\hphantom{\lambda}\bt\ga\delta]}}
\ee
where now the change in the metric can also include variations 
with non-zero
trace. The traceless variations correspond to the addition to 
$\phi$ of an
anti self-dual $4$-form, and the trace variations correspond to 
scaling
$\phi$. Following these results we will now restrict all metric
variations to those which preserve the holonomy group.


\subsection{Metric Variation of $\chi$.}

We may now use what we have learned above to compute the variation
of the self-duality condition obeyed by the field $\chi$, which obeys the
metric dependent condition:
\be
{P_7}\chi =  \chi
\ee
Varying this equation with respect to the metric tensor on $M$ we 
obtain
\be
\delta\chi =  (\delta{P_7})\chi + {P_7}\delta\chi
\ee
Thus
\be
(P_{21})\delta\chi =  (\delta{P_7})\chi.
\ee

The above equation means that in the absence of other relations
relating $\chi$ to the metric tensor, only the variation of $\chi$
into the space of $2$-forms in the ${\bf 21}$ of $Spin(7)$ is 
constrained.
This means that one can consider arbitrary variations of $\chi$ in 
the
${\bf 7}$ of $Spin(7)$. Thus in general, the theory above has a 
continuous
family of energy-momentum tensors. 
However, we find that there is only one choice of this
variation for which the energy-momentum tensor is
BRST exact, which is
\be
({P_7})\delta\chi =  {1\over8}(\delta 
{g^{\al\bt}}g_{\al\bt})\chi.
\ee
The above formula reflects a scaling behaviour of $\chi$ under 
variations
of the metric which have non-zero trace. 

With this choice, the variation of $\chi$ under metric variations
is:
\be
{\delta\chi}_{\al\bt} =  {3\over4}\delta{g^{\rho\sigma}}
g_{\rho[\al}\chi_{\bt]\sigma}
+ {1\over8}\delta{g^{\lambda\rho}}
{\phi_{\lambda\;\;\al\bt}^{\;\;\sigma}}\chi_{\rho\sigma}
+ {1\over8}(\delta{g^{\sigma\tau}}g_{\sigma\tau})\chi_{\al\bt}.
\ee
Using this last formula it is fairly straightforward
to compute the energy-momentum tensor, giving equation $(25)$.

The corresponding formul\ae\ to those above for the 
$7$ dimensional case are as follows:
The stress tensor is
\ba
{T_{\al\bt}} &= & \Tr \{ F^\mu_{\hphantom{\mu}(\al} F_{\beta)\mu} 
+  F^\mu_{\hphantom{\mu}(\al}\tilde{F}_{\beta)\mu}
        -{1\over 4}g_{\alpha\beta}
         F_{\gamma\delta}(F^{\gamma\delta}
      +   \tilde{F}^{\gamma\delta})                       
\nonumber\\
         &+& {3i\over2} 
D_{[\alpha}\psi_{\sigma]}\chi_\beta^{\hphantom{\bt}\sigma}
+ {3i\over2} D_{[\beta}\psi_{\sigma]}\chi_\alpha^{\hphantom{\bt}\sigma}
-{3i\over4}g_{\alpha\beta}D_{\tau}\psi_{\sigma}\chi^{\tau\sigma}\nonumber\\
      &-&D_{(\alpha}\varphi D_{\beta)}\lambda
        +{1\over2}g_{\alpha\beta}D_{\sigma}\varphi 
D^{\sigma}\lambda
    +2iD_{(\alpha}\eta\,\psi_{\beta)}
   - ig_{\alpha\beta}D_{\sigma}\eta\;\psi^{\sigma}       
\nonumber\\
  &-&2i \lambda\psi_\alpha\psi_\beta +
ig_{\alpha\beta}\lambda\psi_\sigma\psi^\sigma
  + {i\over 2}g_{\alpha\beta}\varphi[\eta,\eta] + 
   {1\over 8}g_{\alpha\beta}[\varphi,\lambda]^2 \}.
\ea
This tensor is $Q$-exact:
\be
T_{\al\bt} =  -i\{Q,\la_{\al\bt}\},
\ee
where
\be
\lambda_{\al\bt} =  \Tr\{ {3\over2} F_{(\al}^{\hphantom{\mu}\mu}
                                   \chi_{\bt)\mu} - 
{3\over8}g_{\al\bt}
F_{\ga\delta}\chi^{\ga\delta} + 
                           \psi_{(\al}D_{\bt)}\lambda - 
{1\over2}g_{\al\bt}
                    \psi_\sigma D^\sigma\lambda + 
{1\over4}g_{\al\bt}
                       \eta[\varphi,\lambda] \}.
\ee
The four form has the same variation as in the $8$ dimensional case -
\be
(\delta\phi)_{\al\bt\ga\delta} =  
2\delta 
{g_{[\al\lambda}}{\phi^{\lambda}_{\hphantom{\lambda}\bt\ga\delta]}},
\ee
and the metric variation of the field $\chi_{\al\bt}$ is given by
\be
{\delta\chi}_{\al\bt} =  {2\over3}\delta{g^{\rho\sigma}}
g_{\rho[\al}\chi_{\bt]\sigma}
+ {1\over6}\delta{g^{\lambda\rho}}
{\phi_{\lambda\;\;\;\al\bt}^{\;\;\sigma}}\chi_{\rho\sigma}
+ {3\over28}(\delta{g^{\sigma\tau}}g_{\sigma\tau})\chi_{\al\bt}.
\ee


\section{Invariants}

We may now construct invariants for these theories, following
\cite{Wit}. By these arguments, we deduce that
any correlation function of BRST exact objects
is zero, and any correlation function of BRST invariant objects is
invariant under metric deformations
which preserve the holonomy structure of the manifold $M$.
Defining (these functions are in fact independent of the points at
which the fields are taken, as in \cite{Wit})
\be
W_0 =  {1\over2}\Tr\varphi^2,
\ee
one finds that
\ba
0 &= & i\{Q,W_0\}   \nonumber \\
dW_0 &= &  i\{Q,W_1\}\nonumber \\
dW_1 &= &  i\{Q,W_2\}\nonumber \\
dW_2 &= &  i\{Q,W_3\}\nonumber \\
dW_3 &= &  i\{Q,W_4\}\nonumber \\
dW_4 &= &  0,
\ea
with
\ba
W_1 &= & \Tr(\varphi\psi)\nonumber \\
W_2 &= & \Tr({1\over2}\psi\psi + i\varphi F)\nonumber \\
W_3 &= & \Tr(i\psi F)\nonumber \\
W_4 &= & -{1\over2}\Tr(FF),
\ea
where we understand $\varphi,\psi$ and $F$ to be zero, one and two 
forms
on the manifold $M$ and there is an implicit understanding of a 
wedge
product in all the above formulae. In addition to the above 
quantities one
can consider a further four quantities which can then be used to 
define
correlation functions. These are the following:
\be
W_{i+4} =  {W_i}\wedge\phi
\ee
for $i=1,2,3,4$. These are obtainable from equations 
similar
to those above by replacing $W_0$ with ${W_0}\wedge\phi$, 
since the $Spin(7)$ structure is BRST invariant. 

Now, if $\gamma$ is a $k$ dimensional homology cycle on $M$ then
\be
I(\gamma) =  \int_\gamma W_k,
\ee
is BRST invariant by the above relations, and depends only upon 
the
homology class of $\gamma$ up to BRST exact pieces.
Correlation functions of the $I(\gamma)$ are
then invariant under metric deformations
which preserve the holonomy structure.


\section{Floer Formulation}
In \cite{Wit} the relationship between the four-dimensional 
topological 
field theory and the Floer cohomology groups of three-manifolds
was described. One may ask if there is a relationship between 
our $D$-dimensional theories and some cohomological theory in 
$D-1$ dimensions. In this section we outline 
such a construction for our eight-dimensional theory.
The key to the relationship between Donaldson theory and Floer 
theory
is the existence of a Hamiltonian formulation of the 
four-dimensional
topological field theory on $Y_3 \times R^1$, where $R^1$ 
is the time direction.
In a similar manner, we propose that the Hamiltonian formulation 
of our $D= 8$ theory on $Y_7 \times R^1$ leads to a 
cohomological theory on 
$Y_7$, which in this case is a manifold of $G_2$ holonomy.

Firstly, we will need the identity
\be
D^\al(\lambda_{\al\bt} + U_{\al\bt}) =  0,
\ee
where $\lambda$ is given by eqn. $(27)$, and the antisymmetric
tensor $U_{\al\bt}$ is
\ba
U_{\al\bt} &= & {1\over2}\Tr \{ -(F_{\al\bt} - 
\tilde{F}_{\al\bt})\eta
                 + \phi_{\al\bt}^{\hphantom{\al\bt}\gamma\delta}
                   \psi_\gamma D_\delta\lambda + [\varphi,\lambda]
  \chi_{\al\bt} \nonumber\\       
&& +2 (\tilde{F}_{\al\gamma}\chi_\bt^{\hphantom{\bt}\gamma} - 
\tilde{F}_{\bt\gamma}\chi_\al^{\hphantom{\al}\gamma} ) \}.
\ea
The relatively straightforward
proof of this identity involves the use of the equations of motion 
and the
identities given earlier. Now, following \cite{Wit},
for manifolds $M= Y_7\times R^1$, with $Y_7$ a compact 
seven-dimensional
manifold of $G_2$ holonomy, define
\be
H =  \int d^7x \;T_{00}, \qquad \bar Q =  2 \int d^7x 
\;\lambda^{00}.
\ee
Then
\be \{Q,\bar Q\} =  2H
\ee
and
\be
[H,\bar Q] =  0.
\ee
The proof of this last relation mirrors that in section 4 of 
\cite{Wit},
involving the use of the 
identity given above for the divergence of $\lambda_{\al\bt}$.

The supersymmetry current is given by
\be
J_\mu = \Tr\{-{i\over2}\varphi[\psi_\mu,\lambda] + 
            i(F_{\mu\nu} + \tilde{F}_{\mu\nu})\psi^\nu - i\eta D_\mu\varphi 
         - 2iD^\nu\varphi\,\chi_{\mu\nu} \},
\ee
and is conserved, $D^\mu J_\mu = 0$, using the equations of motion.
Then we have $Q = -i\int_{{Y_7}}J^0$.
Let indices $i,j,...$ run from $1$ to $7$. Define an operation $T$ 
which maps
\ba
& A_i \rightarrow A_, \qquad
         A_0 \rightarrow - A_0, \nonumber\\
& \eta \rightarrow   -\psi_0,  \qquad
       \psi_0 \rightarrow \eta, \nonumber\\
&\varphi \leftrightarrow \lambda, \qquad
 \psi_i \rightarrow 2\chi_{0i}, \nonumber\\
&\chi_{0i} \rightarrow -{1\over2}\psi_i,  \qquad
\chi_{ij} \rightarrow {1\over2}\phi_{ijk0}\psi_k,
\ea
and which maps $t \rightarrow -t$.
Then $T$ maps
\be
Q \rightarrow - \bar Q, \qquad \bar Q \rightarrow Q.
\ee
It is also straightforward to check that the Hamiltonian is
invariant under $T$. Thus, since $Q^2=0$, we have also
\be
\bar Q^2 = 0.
\ee

The operators $Q$ and $\bar Q$ form the basis for the discussion 
of
the Floer invariants.
A similar construction should also work in seven-dimensions where 
the seven-manifold has the structure $Y_6 \times R^1$ and $Y_6$ is 
now
a Calabi-Yau threefold.

\section{Discussion and Conclusion.}

Even though the theories we have discussed here are very much 
based
on the theory in \cite{Wit}, there is a marked difference. Whereas 
the
theory discussed in \cite{Wit} is a topological quantum field 
theory on an
arbitrary $4$-manifold, the theories presented here are only 
``topological''
on manifolds with reduced holonomy groups. This is 
because the $4$-form with which one defines the $D= 4$ instanton 
exists on all
$4$-manifolds, but in higher dimensions a $4$-form which is a 
singlet of
the holonomy group only exists (at least up to $D= 8$) for the 
manifolds
considered here.

The obvious question which springs to mind is - What is special 
about
this particular list of holonomy groups? At least in physics, 
these
holonomy groups are important because manifolds with these 
holonomy groups
give supersymmetric vacua of string theory and $M$-theory 
\cite{chsw,gp,bsa}.
One can then ask if there exists any connection between these 
facts and
the theories we have been discussing. We will now argue that a 
link between
string theory and these theories does indeed exist.

\subsection{Field Theories From Super Yang-Mills Theory
}

(Some of the results of this section have been independently 
obtained in \cite{gtmb}. We are grateful to these authors for discussions.)
Let us first discuss the theory which has been the main focus of 
our paper,
the $D= 8$ theory on manifolds of $Spin(7)$ holonomy. By 
construction, the
theory is supersymmetric with a scalar supercharge $Q$. The theory 
is
not locally supersymmetric, so if there is to be some link with 
some
other theories in eight dimensions it would presumably be another 
globally
supersymmetric theory in eight dimensions. There is only one other 
such theory
in flat space: $D= 8$ super-Yang-Mills theory. With Euclidean 
signature this
theory arises as the effective world-volume theory on a Euclidean
Dirichlet $7$-brane
in Type IIB string theory \cite{db}. In the absence of $D$-branes 
Type IIB
theory is already a supersymmetric theory, so one natural question 
to ask is
what happens when we consider a Euclidean $7$-brane ``wrapped'' 
around a 
manifold of $Spin(7)$ holonomy? 

Let us compute the field content of this curved $7$-brane theory. 
We can
view the curved theory as a ``compactification'' of the $D= 8$ 
super-Yang-Mills
theory in flat space. The $8$-dimensional Euclidean Lorentz group 
is
$SO(8)$. The bosonic field content of the theory consists of a 
gauge field
(in the ${\bf 8_v}$) and two scalars. The sixteen fermions are in 
the
${\bf 8_s}$ and ${\bf 8_c}$. The sixteen supercharges have the 
same
$SO(8)$ labels as the fermions. Under the reduction of $SO(8)$ to 
$Spin(7)$
one gets the following decompositions:
\be
{\bf 8_v \longrightarrow 8}
\ee
\be
{\bf 8_c \longrightarrow 8}
\ee
\be
{\bf 8_s \longrightarrow 7 + 1}
\ee
and the two scalars remain as bosonic scalars. Thus the bosonic
field content of the
theory in curved space consists of two scalars and a gauge field. 
The fermion
field content consists of a scalar, a $1$-form and a field in the 
${\bf 7}$
of $Spin(7)$. This is precisely the field content of the $D= 8$ 
theory we
have been discussing. Furthermore, from the sixteen supercharges 
of the flat
space theory we get one scalar supercharge in the curved space 
theory. This
again is reflected in our $D= 8$ theory. Finally, note that $D= 8$
super Yang-Mills theory is 
essentially a unique theory and thus in considering the theory on 
a curved manifold one should expect a 
unique theory
with the above field content. We therefore claim that the $D= 8$ 
theory we have
discussed in this paper is just super Yang-Mills theory on a 
manifold with
$Spin(7)$ holonomy. Further evidence to support this claim is that 
under the $SO(2)$ R-symmetry of $D= 8$ super Yang-Mills the 
fields 
will have the same ghost numbers as in our theory.
The key point however is that one does not have to twist the 
super-Yang-Mills
theory to arrive at the ``topological'' theory. This contrasts 
with other
topological field theories in four dimensions, however see our 
comments in the 
final subsection below. 

\subsection{D = 7, 6}

One can then go on to consider whether or not some similar 
interpretation 
as that just given exists for the theories we have 
discussed 
on manifolds of $G_2$ and $SU(3)$ holonomy. There is a unique 
super Yang-Mills
theory in $D= 7$ dimensions. Considered as a theory on a manifold 
of $G_2$
holonomy one finds precisely the same field content as we have in 
the
theory with the addition of two scalars: one bosonic and the other 
fermionic.
On closer inspection of the super Yang-Mills action one finds that 
these
``unwanted'' scalars would lead to terms in the curved space 
action which
will be $Q$-invariant. There would also be other terms mixing with 
the
other fields. However, one should also note that the supercharges 
give
rise to two scalar supercharges in the curved space theory, 
whereas the theory
we have been discussing has only one. Setting 
to
zero these unwanted scalar fields may give the field 
theory we
have constructed as a reduction of the ${N_T}= 2$ theory to 
${N_T}= 1$. 
The $7$-dimensional theory could thus be considered as the theory 
obtained
by ``wrapping'' a Euclidean $6$-brane in Type IIA theory around a 
manifold
of $G_2$ holonomy. One is free to argue similarly for the case of 
the $D= 6$
theory on Calabi-Yau threefolds. This theory would then be 
associated with
Euclidean Dirichlet $5$-branes in Type IIB theory on a Calabi-Yau 
threefold.
Combining these results with those of \cite{bsv,gm} gives a
unified picture
in which in all supersymmetric compactifications of string theory, 
D-branes
wrapped around the supersymmetric cycles give ``topological'' 
world-volume theories. Of course, the restriction in the theories we
have discussed here to metric variations which are Ricci flat is
also very natural from the string theory point of view. This is
because non-Ricci flat metrics do not provide a solution to the
low energy field equations in superstring theories.

In fact in the work of \cite{bsv,gm}, the supersymmetric 4-cycles
in 7 or 8 manifolds of exceptional holonomy are shown to be
calibrated submanifolds \cite{hl} where the calibration is the four form
$\phi$ that we have discussed in this paper. The calibrated submanifolds
are unique in each homology class of the manifold and are precisely
the four submanifolds for which $\phi$ is the volume form. These
submanifolds are those about which one can wrap D-branes without breaking all 
supersymmetry. The theory of calibrated submanifolds therefore 
involves an interesting interplay between geometry and topology. 
The theories that we have discussed in this paper 
reflect this interplay and thus should be of importance also 
in the mathematical study of calibrated submanifolds. 

\subsection{Relation to Heterotic and Type I String Theory.}

There is a further possible 
relation to the heterotic and Type I string
theories. In these theories the super-Yang-Mills multiplet
explicitly appears in
the low energy dynamics. Thus when these theories are compactified 
on
manifolds with the holonomy groups we have discussed, a sector of 
the
theory on the internal manifold could plausibly be
describable by the theories we
have presented. This is perhaps not surprising given the relation 
to $D$-branes
we have just discussed. The reason for this is that 
non-perturbative dualities
under which certain Type II and $M$-theory vacua are believed to 
be equivalent
to certain other heterotic/Type I vacua, often map the gauge 
symmetries 
associated with $D$-branes to those in the heterotic and Type I 
theories.
Dualities between heterotic/TypeI/TypeII and $M$-theory 
compactifications
on manifolds with exceptional holonomy have been discussed in 
\cite{gp,bsa}.
We believe that the relationship between the field theories
discussed here and heterotic and Type I strings deserves much 
further study \cite{aol}.

\subsection{A Further Relation With Donaldson-Witten Theory.}

As we discussed above, there is one contrast between the theories
discussed here and that of \cite{Wit}: it is not possible to 
formulate these
theories on arbitrary $D$-dimensional manifolds. Remarkably these
theories still turn out to be \lq\lq topological'', in the sense discussed
above, even though they only exist 
on manifolds
with reduced holonomy groups. One can ask if this property is 
reflected in 
Donaldson-Witten theory. 

We showed in \cite{bm} that the curvature $2$-form for manifolds 
with
holonomy $Spin(7)$, $G_2$, $SU(3)$ and $SU(2)$ satisfies the
instanton equations in higher dimensions. The first three groups 
are the
holonomy groups for the manifolds on which we have defined the 
theories
discussed here. The last group in this list is the holonomy group 
of
hyper-K\"{a}hler manifolds in four dimensions. Given our preceding 
comments
one can then ask, what are the properties of the Donaldson-Witten 
theory
on a hyper-K\"{a}hler manifold? As is well known, the theory in 
\cite{Wit} is
a twisted version of $N= 2$ Super-Yang-Mills theory. However if 
one
considers the theory on a manifold with $SU(2)$ holonomy one finds 
that
the theory after twisting is the same as that before twisting 
\cite{vw}!
This is
due to the fact that the four-dimensional Lorentz group is 
$SU(2)\times SU(2)$, but when the manifold is hyper-K\"{a}hler one of 
the $SU(2)$
factors acts trivially on all fields. 
Thus this is another key property of the four-dimensional theory 
which is
reflected in higher dimensions. The fact that the 
theories discussed above on reduced holonomy manifolds
are topological (in the sense discussed here)
 without twisting stems from the fact these 
manifolds admit spinors which are singlets of the holonomy group.

\bigskip
\bigskip

\noindent{\bf Note Added:} Whilst we were writing this paper, the work 
\cite{bks}
appeared. 
Amongst other results, this paper discusses the gauge-fixing of the action
$S_2$ in a number of cases, and
makes several similar observations to our 
work. The proof that these theories are invariant under metric deformations
preserving the holonomy structure however requires 
the analysis of the BRST exactness
of the stress tensor which we have presented here. In addition
related work with a more mathematical focus has been brought to 
our attention \cite{dt}. 

\newpage

{\centerline {\bf Acknowledgements}}
We would like to thank Jos\'e Figueroa-O'Farrill for
pointing out the source of the error in the initial version of
this paper. We would also like to thank G. Thompson for discussions. 
M.O'L. acknowledges support from EEC Project CHRX-CT93-0340
during the completion of this work.

\end{document}